\documentclass[10pt,letterpaper]{article}
\usepackage{opex3}
\usepackage[ansinew]{inputenc}
\usepackage{amsmath}

\providecommand{\eqref}[1]{(\ref{#1})}

\newcommand{\Pmax}{P_\text{max}}
\newcommand{\Pmin}{P_\text{min}}

\newcommand{\PTC}{P_\text{TC}}
\newcommand{\PAC}{P_\text{AC}}
\newcommand{\PC}{P_\text{C}}
\newcommand{\PA}{P_\text{A}}
\newcommand{\PB}{P_\text{B}}
\newcommand{\XA}{X_\text{A}}
\newcommand{\XB}{X_\text{B}}
\newcommand{\XI}{X_I}
\newcommand{\DA}{D_\text{A}}
\newcommand{\DB}{D_\text{B}}
\newcommand{\DI}{D_I}
\newcommand{\KT}{K_\text{T}}
\renewcommand{\P}{\mathcal{P}}
\newcommand{\PNAB}{P_\text{NAB}}
\newcommand{\PNA}{P_\text{NA}}
\newcommand{\PNB}{P_\text{NB}}
\newcommand{\PNI}{P_{\text{N}I}}
\newcommand{\PNJ}{P_{\text{N}J}}
\newcommand{\TA}{T_\text{A}}
\newcommand{\TB}{T_\text{B}}
\newcommand{\TI}{T_I}

\newcommand{\RA}{R_\text{A}}
\newcommand{\RB}{R_\text{B}}
\newcommand{\RI}{R_I}

\newcommand{\eA}{\eta_\text{A}}
\newcommand{\eB}{\eta_\text{B}}
\newcommand{\tauA}{\tau_\text{A}}
\newcommand{\tauB}{\tau_\text{B}}
\newcommand{\nuF}{\nu_\text{F}}
\newcommand{\nup}{\nu_p}
\newcommand{\nA}{n_\text{A}}
\newcommand{\nB}{n_\text{B}}
\newcommand{\xA}{x_\text{A}}
\newcommand{\xB}{x_\text{B}}
\newcommand{\FS}{F_\text{sys}}
\newcommand{\FSPDC}{F_\text{SPDC}}

\begin{document}

%%%%%%%%%%%%%%%%%% title page information %%%%%%%%%%%%%%%%%%
\title{Simple performance evaluation of pulsed spontaneous parametric down-conversion sources for quantum communications}

\author{Jean-Loup Smirr, Sylvain Guilbaud, Joe Ghalbouni, Robert Frey, Eleni Diamanti, Romain Alléaume and Isabelle Zaquine}

\address{Laboratoire Traitement et Communication de l'Information, Télécom ParisTech, CNRS, Institut Télécom, 46 rue Barrault, 75013 \textsc{Paris}, France}

\email{isabelle.zaquine@telecom-paristech.fr} %% email address is required



\begin{abstract}
Fast characterization of pulsed spontaneous parametric down conversion (SPDC) sources is important for applications in quantum information processing and communications. We propose a simple method to perform this task, which only requires measuring the counts on the two output channels and the coincidences between them, as well as modeling the filter used to reduce the source bandwidth. The proposed method is experimentally tested and used for a complete evaluation of SPDC sources (pair emission probability, total losses, and fidelity) of various bandwidths. This method can find applications in the setting up of SPDC sources and in the continuous verification of the quality of quantum communication links.
\end{abstract}

 % REPLACE WITH CORRECT OCIS CODES FOR YOUR ARTICLE

%%%%%%%%%%%%%%%%%%%%%%% References %%%%%%%%%%%%%%%%%%%%%%%%%

%%%%%%%%%%%%%%%%%%%%%%%%%%  body  %%%%%%%%%%%%%%%%%%%%%%%%%%
\section{Introduction}

Entanglement is a precious resource for many quantum information processing protocols, including quantum key distribution \cite{gisin02}, quantum teleportation \cite{Kim01}, entanglement swapping \cite{Halder07}, quantum relays \cite{Collins05,Aboussan10} quantum memories and repeaters \cite{Briegel98,Duan01,Simon07,Hammerer10}, as well as quantum algorithms \cite{Politi09,Obrien09}. In optics, entangled states are most frequently produced using spontaneous parametric down conversion (SPDC) in nonlinear crystals [12--20].
%\cite{Shih94-Haase09}.
For all the aforementioned applications the fidelity of the entangled states produced by the SPDC source is an essential parameter since it strongly affects the success probability and performance of a specific protocol. Even when systematic imperfections such as unbalanced probabilities between the two components of the entangled state, or residual spectral, temporal and spatial distinguishabilities are corrected, the fidelity of the generated entangled state is still altered by accidental coincidences occurring between photons from two different pairs since there are no longer quantum correlations in this case. This property of SPDC sources has already been addressed in [21--23]. %\cite{Fasel04,Ling09}
However, a means of fast characterization of SPDC sources is, to our knowledge, still missing.

In this paper, we propose a simple method for evaluating the performance (pair emission probability, total losses including detection efficiency, and fidelity) of a quantum link using a pulsed SPDC source, by measuring solely the count probabilities on both output channels and the coincidences between them. Note that these quantities have to be measured in any case in most quantum information processing and communication applications. In the case of fiber quantum communications, the coupling efficiency to the optical fiber can also be deduced from these measurements, provided propagation losses and detection efficiency have been previously measured.The results presented here may be useful in various schemes where entangled photon states are used, especially for setting up high quality sources by standardizing their optimization process. The presented technique may also be useful in entanglement based quantum cryptography to verify in situ and in real time the quality of the quantum link between Alice and Bob.

The paper is organized as follows. Section~2 presents the theoretical foundations of our method. The experimental setup used to test the proposed method is described in Section~3. Section~4 is devoted to the experimental validation of the method, while Section~5 presents results of the measurement of performance of SPDC sources of different bandwidths.

\section{Theoretical analysis}

\noindent In the subsequent analysis, we consider the scheme shown in Fig. 1. In particular, we assume a pulsed pumped collinear quasi-degenerate down conversion in type I (or II) nonlinear crystals. The two photons of the pairs are filtered by the same filter with a bandwidth much larger than that of the pump beam and are coupled to the same optical fiber. The splitting of the photon pairs is ensured by a fiber coupler and photons are finally detected using single-photon avalanche photodiodes. In the following,  our analysis is restricted to this particular case to be consistent with the experimental results presented in Sections 4 and 5, but the procedure can easily be extended to different experimental setups.
%Similar results can be obtained using the presented procedure for other cases with appropriate experimental adjustments.

\begin{figure}[htbp]
	\centering
	\includegraphics[width=.7\textwidth]{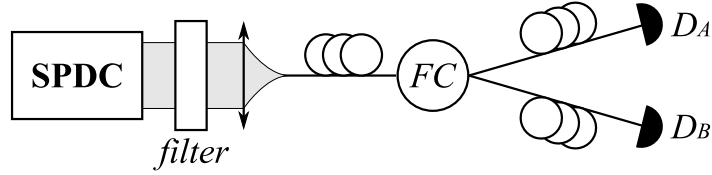}%.jpg
	\caption{Schematic setup of the quantum link. FC: fiber coupler, $\DA$, $\DB$: single-photon avalanche photodiodes. The SPDC crystal is pumped by a pulsed laser.}
\end{figure}

The setup shown in Fig.1 can be idealized as a pulsed SPDC source emitting photon pairs with a peak spectral probability density $p_0$ and two channels exhibiting losses due to filtering, coupling to the fiber, propagation in fibers and detection efficiencies. The splitting of photons between the two channels is necessarily statistical, using the fiber coupler considered here. 
The total transmission on channel $I$ ($I=$A,B) is modeled as the product of a frequency-independent transmission $\XI$ and a frequency-dependent term involving the phase matching condition $G$ and the filter shape $F$.
The transmission $\XI = \RI \TI \eta_I$ takes into account insertion losses of the filter and all other in-line losses $\TI$, the output ratio of the fiber coupler $\RI$ and the quantum efficiency $\eta_I$ of the detector.
The filter transmission $F(\nu-\nu_F)$, which shape is given by an even function $F(\nu)$, has a central frequeny $\nu_F$. 
%The frequency dependence of the transmissions and is described by the product of the maximum transmissions $\XA$ and $\XB$ of channels $A$ and $B$ respectively (at the central frequency $\nuF$ of the filter) and the normalized $G(\nu)$ and $F(\nu)$ transmission functions for the phase matching and filtering processes respectively. 
%Total losses characterized by the transmission $\XI$ in channel $I$ ($I=A,B$) can be decomposed using $\XI = \RI \TI \eta_I$ where $\RI$ is the coupling efficiency to channel $I$,  $\TI$ is the transmission efficiency of path $I$ (including losses due to coupling to the fiber, central frequency transmission of the filter and propagation losses on the considered path), and $\eta_I$ is the quantum efficiency of detector $D_I$. 
Considering a practical SPDC source for quantum communications with a balanced fiber coupler ($\RA \approx \RB \approx 0.5$), low overall losses ($\TA \approx \TB \approx 0.6$), and detector quantum efficiencies $\eA \approx \eB \approx 0.1$, leads to $\XA,\XB \ll 1$.
As the considered SPDC source is pulsed, emitting Gaussian shaped pulses with a half duration $\Delta t$ at $1/e$, the detectors are gated (gates of duration $T$).
The peak spectral probability density $p_0$ is assumed to be small so that the down-conversion probability within the filter bandwidth is low (this corresponds to useful setups in quantum communications).

The various probabilities are experimentally available by measuring counts on detectors $\DA$ and $\DB$ and coincidences between them,  and on the other hand they can be theoretically evaluated from the characteristics of the setup. The derivation of their theoretical expression is detailed in the Appendix, so that only the results that can be useful for the comparison with the experimental data of Section~4 and 5 are presented here.

Let us rewrite and discuss Eqs. \eqref{AppPI}, \eqref{AppPTC}, and \eqref{AppPAC} of the Appendix. The probability $P_I $ of getting a count on detector $D_I$ within the gate duration becomes:
\begin{equation} \label{2PI} 
	P_I= 2 p_0 I_1 \XI \KT + \PNI  \qquad I=\text{A},\text{B}
\end{equation}
where $\PNI$ is the probability of a dark count on detector $\DI$ within the gate duration $T$.

The probability $\PC$ of measuring a coincidence between counts on detectors $\DA$ and $\DB$ within the gate duration $T$ is the sum of three terms :
\begin{equation} \label{eq3} 
	\PC = \PTC+ \PAC+ \PNAB 
\end{equation}

\begin{itemize}
	\item $\PTC$ is the probability of true coincidences due to signal and idler photons of one pair (with respective frequencies $\nu_s$ and $\nu_i$ satisfying $\nu_s+\nu_i=\nu_p$) :
	\begin{equation} \label{2PTC} 
		\PTC= 2 p_0 I_2 \XA \XB \KT  
	\end{equation} 
	\item $\PAC$ is the probability of accidental coincidences between photons of different pairs :
	\begin{equation} \label{2PAC} 
		\PAC= 4 \left( p_0 I_1 \right)^2 \XA \XB \KT^{2} = \left( \PA -\PNA \right) \left( \PB - \PNB \right)
	\end{equation} 
	Note that the expression obtained for $\PAC$ turns out to coincide with the intuitive result for the accidental coincidence probability, i.e. the product of the probabilities of two independent random counts on detectors $\DA,\DB$.

	\item $\PNAB$ is the probability of coincidences related to noise (one down-conversion photon and one noise count, or two noise counts) :
	\begin{equation} \label{eq8} 
		\PNAB  =\left(\PA -\PNA  \right) \PNB + \left(\PB -\PNB \right) \PNA + \PNA  \PNB 
	\end{equation}
\end{itemize}

\begin{figure}[htbp]
	\centering
	\includegraphics[width=\textwidth]{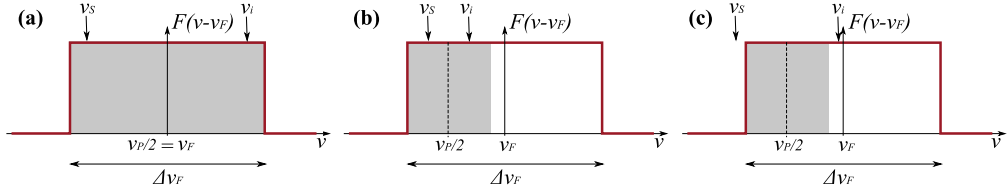}%.pdf
	\caption{Principle of the evaluation of $I_2$ for the simple case of a rectangular filter function. The shaded area is proportional to the effective value of $I_2$. When the filter is centered at the degeneracy frequency (zero detuning), $I_2$ is maximum (a). When detuning is non-zero, $I_2$ is reduced : some photons within the filter bandwidth have their twins transmitted (b), while this is not the case for others (c).\label{I2}}
\end{figure}

In these equations $\KT$ describes the time dependence:
\begin{eqnarray}
		\KT &=& \int_{-T/2}^{+T/2} e^{-\frac{t^2}{\Delta t^2}} dt \Big/ \int_{-\infty }^{+\infty} e^{-\frac{t^2}{\Delta t^2}} dt \label{2KT}
\end{eqnarray}
Assuming that the phase matching function $G \equiv 1$ (as will be justified in section~4 for our setup), the terms $I_1$, $I_2$ take into account the filter shape in the following way:
	\begin{eqnarray}
	I_1 &=& \int_{-\infty}^{+\infty} F(\nu-\nu_F) d\nu   \\ %G(\nu) 
	I_2 &=& \int_{-\infty }^{+\infty}  F(\nu-\nu_F) F(\nu_p-\nu-\nu_F) d\nu =F*F(\nu_p-2\nu_F) \label{eq5} 
\end{eqnarray}
While the whole bandwidth contributes to single counts ($I_1$), the spectral transmission function for coincidences $I_2$ is the auto-convolution of the filter shape. Because $F$ is an even function, $I_2$ is maximum for a zero detuning between the degeneracy frequency and the central frequency of the filter ($\nup/2=\nuF$). %The single count rate,  is independent of this detuning. 
This is due to the fact that symmetric filtering around the degeneracy frequency optimally preserves the correlations between signal and idler photons.
The decrease of $I_2$ when $\nup/2 \neq \nuF$ is illustrated in Fig. \ref{I2} for the case of a rectangular filter. %Its value is proportional to the shaded area, that appears clearly smaller when when $\nup/2 \neq \nuF$ (left and center) than when $\nup/2 = \nuF$ (right).Indeed, as $\nu_s$ and $\nu_i$ are symmetrical with respect to $\nup/2$ for SPDC pairs, when $\nu_i$ is in the unshaded region, $F(\nu_s)=0$ and no coincidence can be observed.
%In other words,  the spectral transmission function for coincidences ($I_2$) is the auto-convolution of the filter transmission $F*F(\nup-2\nuF)$. As such, it is maximum for a zero detuning between the filter central frequency and the degeneracy frequency, while the single count rate is independent of this detuning. 

%When $\nup/2 \neq \nuF$, $I_2$ decreases : this is illustrated in Fig. 2 for the case of a rectangular filter where the value of $I_2$ for $\nup/2 \neq \nuF$ (left and center), proportional to the shaded area, appears clearly smaller than its value for $\nup/2 = \nuF$ (right).

The incoherent nature of the accidental coincidences gives rise to quadratic dependence of $\PAC$ on $I_1$ and $\KT$ (Eq. \eqref{2PAC}) for the spectral filtering and temporal gating. On the other hand, $\PTC$ (Eq. \eqref{2PTC}) is proportional to $I_2$ and $\KT$ because of spectral and temporal correlations within a pair.

Based on the equations given above, it is now straightforward to derive some important figures of merit for the SPDC source, in particular the pair emission probability, total losses and a measure of quality that we will define below.

Dividing Eq. \eqref{2PAC} by Eq. \eqref{2PTC} and rearranging the terms, we obtain the probability $p_0 I_1$ of pair generation within the filter bandwidth:
\begin{equation} \label{eq11} 
	p_0 I_1 =\frac{I_2 }{2I_1 \KT } \frac{\PAC }{\PTC}=\frac{I_2 }{2I_1 \KT } \frac{\left(\PA -\PNA \right) \left(\PB -\PNB \right)}{\left(\PC -\PAC -\PNAB \right)}  
\end{equation}
where $\PNAB$ and $\PAC$ are given by Eqs. \eqref{eq8} and \eqref{2PAC}, respectively. 
The global transmission efficiency $\XI$ is derived from Eqs. \eqref{2PI} and \eqref{2PTC} 
\begin{equation}\label{eq2XI}
	\XI =\frac{I_1 }{I_2 } \frac{\PTC}{P_J -\PNJ } =\frac{I_1 }{I_2 } \frac{\left(\PC -\PAC -\PNAB \right)}{P_J -\PNJ } \qquad (J = \text{B,A} \text{ for } I = \text{A,B}) 
\end{equation}

Note that the temporal correction term $\KT$, which depends on the relative duration of the detection gate and the pump pulse, and could have been interpreted as a loss, does not appear in the expression of the total losses. This is a consequence of the aforementionned strong temporal correlations between twin photons. It is important to note that $p_0I_1$ and $\XI$ are easily obtained from the measurements of $\PA$, $\PB$, $\PC$ provided preliminary calculations of $I_2/I_1$ and $\KT$, and measurements of $\PNA$ and $\PNB$, have been performed. These preliminary calculations and measurements can be done once and for all before the source characterization is started.

Finally, an important property for SPDC sources used in quantum communication systems is the correlation between the counts on channels A and B. %Such a measure will hereafter be designated ``system fidelity'' $\FS$.
The most general measurement of an entangled state is usually performed using quantum state tomography \cite{James01}. However, in entangled state based communications, Bell type measurements are preferred \cite{Ekert91}. In such a case, the quality of a SPDC source is often evaluated from the measurement of coincidence rates as a function of a free parameter. For instance, this parameter can be the relative angle between polarization analyzers of channel A and B in the case of polarization entanglement [12--20]
%\cite{Shih94,Haase09}
or the relative phase between the Franson interferometers of channel A and B in the case of time bin entanglement \cite{Marcikic02}.

The evaluation is made quantitative through the measurement of the visibility (or contrast) of the two-photon interference fringes $V=(\Pmax-\Pmin)/(\Pmax+\Pmin)$ where $\Pmax$ and $\Pmin$ are respectively the maximum and minimum coincidence rates (or probabilities) that can be obtained when varying the free parameter. Practical quantum communication devices require maximally entangled states and no decoherence due to spectral, spatial or temporal distinguishabilities. In this case, the source quality is still altered because of multiple pairs generation (double pairs being dominant thus only considered in this paper). The direct measurement of this visibility necessarily involves detectors so that a system fidelity $\FS$ can be defined as a visibility where $\Pmin=\PAC+\PNAB$ and $\Pmax=\PTC+\PAC+\PNAB$.  :
\begin{equation} \label{eq13} 
	\FS = \frac{1}{1+2\frac{\PAC +\PNAB }{\PTC } }
\end{equation} 
$\PAC$ and $\PNAB$ are given by Eqs. \eqref{2PAC} and \eqref{eq8}, respectively. This fidelity constitutes an upper bound for Bell type measurements that can only be reached when all imperfections other than the unavoidable presence of multiple pairs are corrected.
If the source is going to be part of a more complex quantum communication system, including for instance a quantum memory, it can be useful to define the intrinsic source fidelity $\FSPDC$ (that does not include detector noise):
\begin{equation} \label{eq14} 
	\FSPDC =\frac{1}{1+2\frac{\PAC}{\PTC}}  
\end{equation} 
Using Eqs. \eqref{eq3} and \eqref{2PAC}, $\FSPDC$ can be determined from measurements.
Let us stress the fact that the only prerequisite knowledge to obtain these fidelities is the detector noise.
Eq. \eqref{eq14}  can also be written as
\begin{equation} \label{eq15} 
	\FSPDC =\frac{1}{1+4 (p_0 I_1) \KT \frac{I_1}{I_2} }  
\end{equation} 
which shows that the fidelity is reduced at high pair production probabilities and high $I_1/I_2$ ratios.

From Eqs. \eqref{eq11} and \eqref{eq2XI} it is clear that determining  $p_0 I_1$ and $\XI$ requires the knowledge of the ratio $I_1/I_2$, and therefore the knowledge of the filter shape $F(\nu)$.

\begin{table}[htbp]
	\centering\begin{tabular}{cccc} \hline 
	Case & \phantom{{\Huge $p_0$}} Type \phantom{{\Huge $p_0$}}            & Filter Bandwidth (GHz) & $I_1/{I_2}_\text{max}$ \\ \hline 
	\{1\}& Rectangular      & any                    & 1 \\
	\{2\}& Triangular       & any                    & 1.50 \\
	\{3\}& Gaussian         & any                    & 1.41 \\
	\{4\}& DWDM             & 73                     & 1.14 \\
	\{5\}& DWDM + FP        & 1.63                   & 2.09 \vspace{3pt} \\  \hline 
	\end{tabular}
	\caption{Filter bandwidth and $I_1/{I_2}_\text{max}$ for various filters where ${I_2}_\text{max}=I_2(\nup/2-\nuF=0)$. DWDM stands for Dense Wavelength Division Multiplexing add/drop filter, FP stands for Fabry-Pérot etalon. }
\end{table}

Table 1 presents calculated results obtained for various filters at zero detuning between their central frequency $\nuF$ and the degeneracy frequency $\nu_p/2$. The third column of Table 1 gives the bandwidth of the filter. The ratio $I_1/I_{2_\text{max}}$ listed in the fourth column of the table gives a direct quantification of the influence of the filter shape on the fidelity of the SPDC source (see Eq. \eqref{eq15}). The best possible case is the rectangular filter  \{1\} for which $I_1/I_{2_\text{max}}=1$ irrespective of the bandwidth. The theoretical triangular and Gaussian filters considered in cases \{2\} and \{3\} are less attractive. Case \{4\} and \{5\} concern practical filters used experimentally (see Sections 4 and 5). It is important to note the very good quality of the apodized commercially available DWDM filter that provides a factor $I_1/I_{2_\text{max}}= 1.14$, which is close to the optimal unity value. It is also important to remark that small bandwidths can be obtained with a penalty of only about two for the ratio $I_1/I_2$ when using Fabry-Pérot devices. Further calculations show that much smaller bandwidths can be obtained by cascading FP etalons with no supplementary penalty as far as $I_1/I_2$ is concerned. This result is important in view of designing SPDC sources with very small linewidths, compatible with quantum memories \cite{Goldner09}.

\section{Experimental setup}

\begin{figure}[htbp]
	\centering
	\includegraphics[width=.8\textwidth]{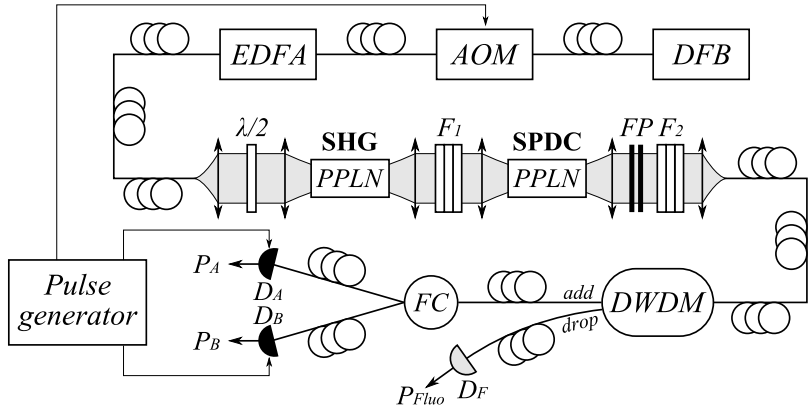}%.jpg
	\caption{Experimental setup. See text for explanation of acronyms.\label{exp}}
\end{figure}

The experimental setup used to verify the theoretical calculations presented in Section~2 and to evaluate the performance of SPDC sources of different bandwidths is depicted in Fig. \ref{exp}. Nearly degenerate SPDC at 1564 nm was performed using a pulsed pump beam operating at 782 nm. This pump beam was obtained starting from a 40 mW power CW DFB laser operating at 1564 nm. The CW laser was modulated at a frequency of 2 MHz using an acousto-optic modulator (AOM) in order to obtain Fourier-transform-limited pulses. The pulsed signal was then amplified up to a 5W mean power by an erbium-doped fiber amplifier (EDFA). The high peak power (100 W) of the 25 ns duration pulses delivered by the EDFA allowed obtaining a high mean power (1.3 W) pump beam at 782 nm through second harmonic generation in a 2 cm long periodically poled lithium niobate (PPLN) crystal. After total elimination of the remaining 1564 nm beam by a set of frequency filters $F_1$ the pump beam was focused into a second 2 cm long PPLN crystal to produce SPDC around degeneracy at 1564 nm. After filtering of spurious light by the filter set $F_2$ photon pairs delivered by SPDC were collected and focused into an antireflection coated monomode optical fiber. The large bandwidth ($\approx 100$ nm) of the emitted SPDC photon pairs was reduced to approximately 70 GHz (0.57 nm) using a commercially available fiber dense wavelength division multiplexing (DWDM) add/drop filter. The measurement of the rejected part of the spectrum (detector $D_F$) can be used  to estimate the total fluorescence power. The photons of each pair were separated (with a 50\% efficiency) by a 50\%-50\% fiber coupler and detected by InGaAs avalanche photodiodes operated in gated mode using the pulse generator driving the AOM for detector synchronization.

The probabilities per pulse $\PA$,  $\PB$ and $\PC$, which are the relevant parameters for pulsed SPDC sources, are obtained by dividing the single and coincidence count rates (per second) by the repetition rate of the SPDC source.

Additional filtering down to 1.63 GHz was also performed by inserting a 2 mm thick Fabry Pérot (FP) etalon in front of filter $F_2$. Note that thanks to the high power achieved at 782 nm, lower bandwidth SPDC sources could also be obtained using additional FP etalons. This would allow reducing the bandwidth enough so as to be compatible with quantum memories that can require very narrow ($< 100$ MHz) linewidths.

\section{Validity of the procedure}

The non collinear phase matching assumption used in the calculations to allow $G \equiv 1$ has been validated by a measurement of the fluorescence spectrum collected in the optical fiber when no filtering is made. The shape of the spectrum can be considered Gaussian with a full width at half maximum of more than a hundred nanometers. As the considered filter bandwidths are smaller than 1 nm, the approximation $G \equiv 1$ used in Section~2 is satisfied with a relative precision better than $10^{-6}$.

The fidelity factors $\FS$ and $\FSPDC$ given by Eqs. \eqref{eq13} and \eqref{eq14} respectively require no parameters for their calculation other than count rate measurements, thus the obtained values can be considered fully reliable. On the other hand, the relevance of the evaluation of $p_0 I_1$ and $\XI$ depends on the validity of the preliminary calculations of $I_1$ and $I_2$. Therefore, before applying results of Section~2 to obtain a simple evaluation of performance of SPDC sources of various bandwidths, preliminary experiments were performed to verify the validity of using computations of $I_1$ and $I_2$ to infer reliable values for $p_0 I_1$ and $\XI$. To this end, $I_1$ and $I_2$ were both calculated and derived from measurements made for different values of $\nu_p/2$, by changing the temperature of the DFB laser. This operation was performed in the cases of a DWDM filter of bandwidth $\Delta \nuF=$ 73 GHz and a filter set of bandwidth $\Delta \nuF = 1.63$ GHz composed of the same DWDM filter and a solid FP etalon (free spectral range = 50.0 GHz and finesse = 31.5 calculated from the measured thickness and mirror reflectivity of the FP etalon). 
Calculations of $I_1$ and $I_2$ were performed using the trapezoidal shape for the DWDM filter (case \{4\} of Table 1) and its product with the Airy function centered on $\nuF$ for the FP etalon (case \{5\} of Table 1). To compare this calculated value of $I_2(\nup)$ to an experimentally derived one, we used Eq. \eqref{2PTC}. It predicts that the frequency dependence of $\PTC$ should coincide with that of $I_2$. 

\begin{figure}[htbp]
	\centering
	\includegraphics[width=.73\textwidth]{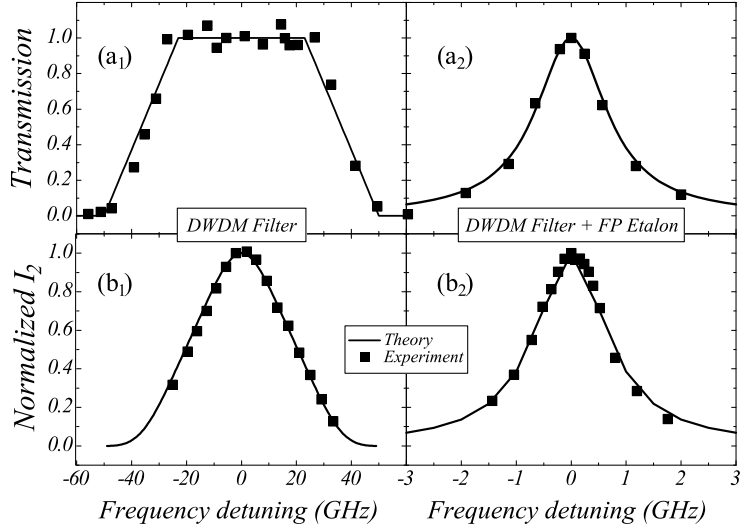}%.pdf
	\caption{Comparison of the theoretical and experimental transmission (a) and normalized value of $I_2$ (b) plotted as a function of frequency detuning with respect to the filter center frequency $\nuF$, for two different filtering devices: a DWDM filter \{4\} and a DWDM filter plus a FP etalon \{5\}.\label{valid}}
\end{figure}

Figs. \ref{valid}.a$_1$ and a$_2$ show calculated and measured transmission spectra for $\Delta \nuF= 73$ GHz and $\Delta \nuF = 1.63$ GHz respectively. The very good agreement observed validates our choice of the theoretical filter shape $F(\nu)$.  In Figs. 4.b$_1$ and b$_2$, normalized spectra of $I_2/I_{2_\text{max}}$ and $\PTC/P_{\text{TC}_\text{max}}$ can be compared. The excellent agreement between experimental and theoretical results confirms the validity of the analysis of Section~2 and hence the importance of using a filter very well centered on the degenerate frequency $\nup/2$ of the SPDC process in order to obtain the best possible performance of the photon pair source.

\section{Evaluation of source performance}

The results of Section~2 were used to evaluate the performance of SPDC sources. The two filters of different bandwidths described in Section~4 were tested : a fiber DWDM filter ($\Delta \nuF=$ 73 GHz, case $\{4\}$ of Table 1) and a set composed of the fiber DWDM filter and a free-space FP etalon ($\Delta \nuF=$ 1.63 GHz, case $\{5\}$ of Table 1).

Preliminary measurements of the detector characteristics were performed giving $\eA$ = 0.080 and $\eB$ = 0.076 for the quantum efficiencies, and $\PNA = 1.9 \times 10^{-4}$ and $\PNB = 1.5 \times 10^{-4}$ for the dark count probabilities of detectors $\DA$ and $\DB$, respectively. For each filter we measured the total fluorescence mean power $P_{Fluo}$ dropped by the DWDM filter, the single count probabilities $\PA$ and $\PB$ for detectors $\DA$ and $\DB$, and the coincidence probability $\PC$ between the detector counts. The probability of pair generation within the filter bandwidth $p_0 I_1$ was calculated using Eq. \eqref{eq11} for $\nup/2=\nuF$ as well as the value of $I_1/I_{2_\text{max}}$ given in Table 1 for the filter. We also used the value $\KT = 0.75$ obtained for a detection window $T = 20$ ns and a full width at half maximum $2 \sqrt{2 \ln 2} \, \Delta t = 20.3$ ns for the Gaussian pump pulse at 782 nm (we also verified that signal and idler pulses had the same duration as the pump pulse). 

\begin{figure}[htbp]
	\centering
	\includegraphics[angle=0,width=.98\textwidth]{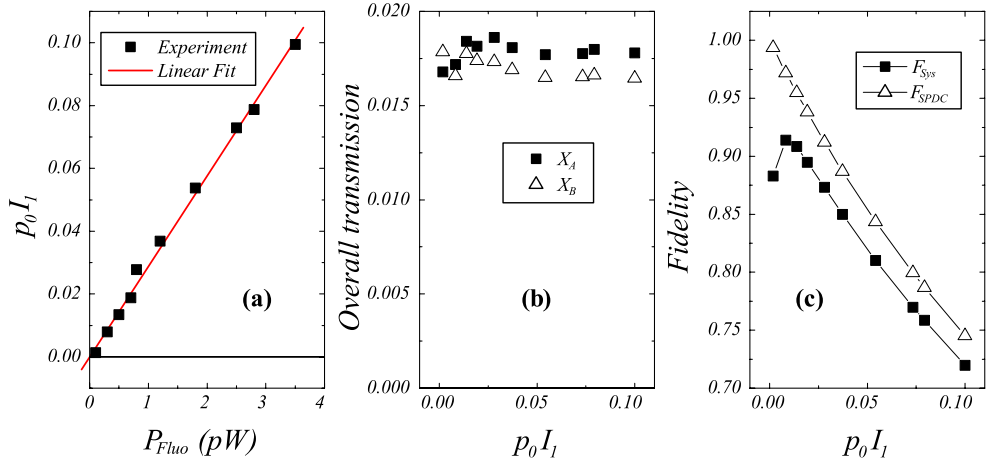}%.pdf
	\caption{Performance of our SPDC source using a DWDM filter.\label{results}}
\end{figure}

The probability $p_0 I_1$ is plotted in Fig. \ref{results}.a as a function of the total fluorescence mean power dropped by the DWDM filter. The proportionality between these two quantities is an additional proof of the validity of our analysis.
The overall transmission, experimentally determined using Eq. \eqref{eq2XI}, is plotted in Fig. 5.b for the DWDM filtering device. As expected, the coupling efficiency does not depend on $p_0 I_1$. Using the values of Fig. 5.b, we find $\XA= 0.0178 \pm 0.001$ and $\XB = 0.0170 \pm 0.002$ for the DWDM filter alone. Similar measurements made with the DWDM plus Fabry-Pérot filters set give $\XA = 0.0150 \pm 0.001$ and $\XB = 0.0141 \pm 0.001$. Note the low standard deviation: this indicates that a precise control could be operated on the quality of a specific quantum link.

Knowing that $\XI = \RI \TI \eta_I$ allows the derivation of $\TI$ provided $\RI$ and $\eta_I$ have been previously determined by auxiliary measurements. $\TI$ is in fact the product of propagation and filter transmission $\tau_I$ and fiber coupling efficiency $C_\text{F}$. The preliminary measurement of $\tau_I$ by reverse propagation through the system (in particular, by entering a fiber laser source at frequency $\nup$/2 into the output fiber from detectors $\DA$ and $\DB$) allows then a precise derivation of the coupling efficiency of photon pairs into the optical fiber, which is very difficult to determine otherwise.

The coupling efficiency $C_\text{F}$ to the optical fiber was derived in the case of the DWDM filter using the measured values $\RA \tauA$ = 0.301 and $\RB \tauB$ = 0.308 . Its high value ($C_\text{F}=0.74$) evidently demonstrates the high quality of the coupling of photon pairs generated by SPDC in our setup.

In the case of the DWDM plus FP filter set, we derived the FP transmission from the comparison between the mean values of $\XA$ and $\XB$ obtained with the two filtering set-ups:  the value $\tau_\text{FP} = 0.84 \pm 0.01$ was obtained. This value is somewhat lower than the value of $0.99$ obtained from direct transmission measurements performed using counter propagation with an auxiliary laser. This could be due to the fact that the Fabry-Pérot transmission not only depends on frequency but also on direction. The non-perfect-collinear nature of the phase matching conditions induces a fluorescence beam that has a different profile than the laser beam used to measure the FP transmission. This is not taken into account in the evaluation of $I_2$, thus the value of $I_1/{I_2}_\text{max}$ given in Table 1 could be slightly underestimated.

%This could be due to the dual dependence of the Fabry-Pérot transmission with frequency and direction that is not taken into account in the evaluation of $I_2$: due to non collinear phase-matching, the value of $I_1/{I_2}_\text{max}$ given in Table 1 could be underestimated.

Finally, the system and source fidelity factors $\FS$ and $\FSPDC$ calculated using Eqs. \eqref{eq13} and \eqref{eq14} are plotted in Fig 5.c in the case of the DWDM filter (similar results were obtained with the DWDM plus Fabry-Pérot etalon filtering set) as a function of $p_0 I_1$. As expected, $\FSPDC$ is always greater than $\FS$, and the difference increases for smaller pair generation rate due to the increasing importance of electronic noise. Note that the system fidelity cannot be higher than 92\% due to the noise of the avalanche photodiode detectors. However, the use of superconducting detectors would greatly improve the maximum fidelity \cite{Hadfield06}. In order to observe a Bell inequality violation, the fidelity must be greater than $1/\sqrt{2}$. This limits the pair generation probability to about 10\% in the case of the 73 GHz bandwidth. For the 1.63 GHz bandwidth offered by our other filtering system, the maximum generation probability decreases to 5\%.

\section{Conclusion}

The assessment of the quality of a SPDC source, or an entire link including such a source, used for applications in quantum communications is difficult when only the count and coincidence rates are known. We have shown, however, that a reliable evaluation of the quantum source (or link) performance, including the pair generation probability, overall losses, and fidelity, is simple in the case of pulsed sources emitting maximally entangled states when taking into account the transmission spectrum of the used filter. The method has been checked experimentally and has been successfully applied to the cases of two SPDC sources of different bandwidths.

It is important to note that our analysis is restricted to the case when the occurrence of accidental coincidences is the only limitation to the source quality, unbalanced probabilities of the two-photon components of the entangled state and spectral, temporal, and spatial distinguishabilities being already corrected. This restriction is justified, as these problems have to be corrected anyway in SPDC sources for quantum communications or information processing. This means that our analysis may be extended to other SPDC sources including non collinear down conversion, provided these defects have been previously corrected. The only requirement concerns the pulsed nature of the SPDC source. 
The results of our study could be qualitatively applied to the case of continuous wave SPDC sources for which counts and coincidence rates are the relevant parameters. Indeed, for a detector with a dead time $\Delta T$ that is much larger than the inverse of the filter bandwidth, the relevant time duration corresponding to the inverse repetition rate of the pulsed source is the minimum time between two successive counts, i.e. the dead time of the detectors  $\Delta T$. Our results can then be applied, using probabilities approximated by the product of the measured rates and the dead time $\Delta T$. From a quantitative viewpoint,  the complete theory remains nevertheless to be done. It should be noted that if free running SPDC sources can be safely used in point to point quantum communications, synchronisable pulsed SPDC sources are the most suitable devices for future applications in quantum information processing and communication networks. The procedure described in this paper can then find applications in these devices, in particular for the setting up of the SPDC sources or the fast verification of the quality of the quantum links using SPDC sources.

%This procedure can then find applications in quantum information processing and communications, in particular for the setting up of SPDC sources or the real time checking of the quality of quantum links using SPDC sources.

\section{Acknowledgements}

The authors acknowledge financial support from the Agence Nationale de la Recherche through the ``e-QUANET'' (ANR-09-BLAN-0333-01) and ``FREQUENCY'' (ANR-09-BLAN-0410-CSD1) projects, the Regional Council IDF and the Institut Telecom.

\appendix

\section{Appendix : theoretical evaluation of count rates}

The evaluation of various properties of pulsed SPDC sources (statistical or deterministic splitting of the down-converted photons, unbalanced filtering between channels, partial coherence of multiple pairs within a pulse, ...) will be addressed in a forthcoming paper. Here, the analysis is restricted to the case that has experimentally been investigated : a pulsed SPDC source with a common filter inserted before splitting photons into two paths. Double pairs within a pump pulse, when they occur, are mutually incoherent, that is, they originate from two independent down-conversion processes.

Let $N$ photon pairs be generated by degenerate collinear SPDC in the crystal during a pump pulse, and $\xA$ and $\xB$ be the total transmissions  on channel A and B respectively. The probability of getting $\nA$ and $\nB$ photons on channel A and B respectively, using statistical splitting, is given by:
\begin{equation}
	\P_N(\nA,\nB) = C_{2N}^{2N-\nA-\nB}(1-\xA-\xB)^{2N-\nA-\nB} \, C_{\nA+\nB}^{\nA}\xA^{\nA}\xB^{\nB}
\end{equation}
where $C_a^b=a!/(b!(a-b)!)$. Indeed, in such a case $2N-\nA-\nB$ photons are lost, each with a probability $1-\xA-\xB$ and the detectable photons are distributed between channels A and B with probabilities $\xA^{\nA}$ and $\xB^{\nB}$ respectively. 
The factors $C_{2N}^{2N-\nA-\nB}$ and $C_{\nA+\nB}^{\nA}$ evaluate the number of possible cases corresponding to ${2N-\nA-\nB}$ lost photons and $\nA$ detectable photons on channel A respectively.
The probability of one count on channel A for instance is proportional to the probability of getting at least  one photon on channel A when one pair is produced $\P_1(\nA\geq1,\nB)$:
\begin{equation}\label{AppP1}
\P_1(\nA\geq1,\nB)=\P_1(2,0)+\P_1(1,1)+\P_1(1,0)=\xA(2-\xA)
\end{equation}
%In the setup considered in Fig. 1, the transmission $x_I$ in channel $I$ ($I=A,B$) can be decomposed as $x_I = \XI F(\nu)=\RI \TI \eta_IF(\nu)G(\nu)$ where $\RI$ is the coupling efficiency to channel $I$,  $\TI$ is the transmission efficiency of path $I$ (including losses due to coupling to the fiber, central frequency transmission of the filter and propagation losses on the considered path), $\eta_I$ is the quantum efficiency of detector $D_I$ and $F(\nu)$ is the normalized filter response. Considering a practical SPDC source for quantum communications with a balanced fiber coupler ($\RA \approx \RB \approx 0.5$), low overall losses ($\TA \approx \TB \approx 0.6$), and detector quantum efficiencies $\eA \approx \eB \approx 0.1$, leads to $\XA,\XB \ll 1$ and hence $\xA,\xB \ll 1$ . 
As explained in the main text, in practical SPDC sources for quantum communications, $\xA, \xB\ll 1$. In this case:
\begin{equation}
\P_1(\nA\geq1,\nB)\simeq2\xA
\end{equation}
The probability of getting one count on channel $I$ is therefore given by:
\begin{equation}\label{AppPI}
	P_I =2p_0 \KT \XI \int _{-\infty}^{+\infty} F (\nu-\nu_F)G(\nu)d\nu + \PNI=2p_0 \KT \XI I_1+ \PNI
\end{equation}
where $\XI$, $F$ and $G$ have been defined in the main text and $I_1=\int _{-\infty}^{+\infty} F (\nu-\nu_F)G(\nu)d\nu $. The constant $\KT =\int_{-T/2}^{+T/2} I_p(t) dt / \int_{-\infty}^{+\infty } I_p(t )dt$, where $I_p$ is the pump intensity, is included in Eq. \eqref{AppPI} to account for photon loss due to the detection time window and $\PNI$ is the probability  of registering a dark count on detector $D_I$.

The probability $\PC$ of measuring a coincidence between counts on detectors $\DA$ and $\DB$ comes from the contributions of true and accidental coincidences as well as coincidences related to noise:
\begin{equation} \label{AppPC} 
	\PC = \PTC + \PAC + \PNAB  
\end{equation}

$\PTC$ is the probability of registering simultaneous counts on detectors $\DA$ and $\DB$ due to the two photons of a single pair and is calculated using the probability of properly getting one photon of the pair on each channel $ \P_1(1,1)=2\xA\xB$:
\begin{equation} \label{AppPTC} 
	\PTC = 2 p_0 \KT  \XA \XB \int _{-\infty}^{+\infty} F(\nu-\nu_F)G(\nu)F(\nu_p-\nu-\nu_F)G(\nu_p-\nu)d\nu=2 p_0 \KT  \XA \XB I_2
\end{equation} 
where $I_2$ measures the impact of filtering and phase matching on the coincidence probability of the source and is given by: $I_2=\int _{-\infty}^{+\infty} F(\nu-\nu_F)G(\nu)F(\nu_p-\nu-\nu_F)G(\nu_p-\nu)d\nu$.
Note that $\PTC$ is proportional to $\KT$ and not to $\KT^2$ since the two photons of a pair are emitted simultaneously when the signal and idler bandwidths are much larger than the pump bandwidth.

The second term in Eq. \eqref{AppPC} is the probability of registering simultaneous counts on detectors $\DA$ and $\DB$ due to two signal (or idler) photons of two different pairs: it is proportional to the probability of getting at least one photon on each channel when two pairs have been produced simultaneaously $\P_2(\nA\geq1,\nB\geq1)$ 
\begin{eqnarray} 
\P_2(\nA\geq 1,\nB\geq 1)&=&\left[ \P_2(3,1) + \P_2(1,3) + \P_2(2,2) + \P_2(2,1) + \P_2(1,2) + \P_2(1,1)\right] \nonumber\\
&=&\left[6-6\left(\xA +\xB \right)+2\left(\xA^{2} +\xB^{2} \right)+3\xA \xB \right] \xA \xB \simeq 6\xA \xB
\end{eqnarray}
\begin{eqnarray} 
	\PAC =\frac{2}{3} p_0^2 \KT^2 6\left[\XA I_1\right]  \left[\XB I_1\right] =4p_0^2 \KT^2 \XA\XB I_1^2\label{AppPAC}
\end{eqnarray}

The factor $2/3$ is the proportion of coincidences due to double pairs that effectively involve photons of different pairs.

The third and final term in Eq. \eqref{AppPC}:
\begin{equation} \label{AppPNAB} 
	\PNAB = \left( \PA -\PNA \right) \PNB + \left( \PB -\PNB \right) \PNA +\PNA \PNB  
\end{equation} 
is the probability of registering simultaneous counts on detectors $\DA$ and $\DB$ due to one down-conversion photon on one detector and one dark count on the other one, or dark counts on both detectors.

These equations are discussed in the main text, where they are also used to derive the performance of the source.

\end{document}